\documentclass[twoside]{article}
\usepackage{graphicx}
\usepackage{fancyhdr}
\usepackage{multicol}
\usepackage{styl-ap}

\begin{document}
\title{New insights from inside-out Doppler tomography}
\author{Enrico J. Kotze\work{1,2}, Stephen B. Potter\work{1}}
\workplace{South African Astronomical Observatory, PO Box 9, Observatory 7935,
Cape Town, South Africa
\next
Astrophysics, Cosmology and Gravity Centre (ACGC), Department of Astronomy,
University of Cape Town, Private Bag X3, Rondebosch 7701, South Africa}
\mainauthor{ejk@saao.ac.za}
\maketitle

\begin{abstract}
We present preliminary results from our investigation into using an
``inside-out'' velocity space for creating a Doppler tomogram.
The aim is to transpose the inverted appearance of the Cartesian velocity
space used in normal Doppler tomography.
In a comparison between normal and inside-out Doppler tomograms of
cataclysmic variables, we show that the inside-out velocity space has the
potential to produce new insights into the accretion dynamics in these
systems.
\end{abstract}


\begin{multicols}{2}
\section{Introduction}
Cataclysmic variables (CVs) are quintessential stellar objects for the study
of mass transfer, accretion flows and accretion discs.
The typical interacting system consists of a secondary low-mass, red dwarf
star which is filling its Roche lobe and is transferring mass via the inner
Lagrangian point ($L_{1}$) onto the primary white dwarf star (see Warner 1995
for a comprehensive review).
Doppler tomography, as introduced by Marsh \& Horne (1988), is aimed at
rendering the information locked-up in phased-resolved spectra of a CV into
a two-dimensional map of the binary components in velocity space (Doppler
tomogram).

\section{Doppler tomography}

\subsection{Spatial coordinates}

Fig.~1 shows a model CV with an accretion disc in a Cartesian spatial
coordinate frame that co-rotates with the system.
As described by Marsh \& Horne (1988), this two-dimesional frame has its
origin at the centre of mass of the system [marked with a plus (+)], the
X-axis along the line connecting the centres of mass of the primary and
secondary [marked with crosses ($\times$)], and the Y-axis parallel to the
velocity vector of the secondary.
The orbital motion is assumed to be counter-clockwise around the centre of
mass of the system.
The model CV assumes the following system parameters:
inclination $i=87^{\circ}$; mass of the primary $M_{1}=0.8$; mass ratio
$q=M_{2}/M_{1}=0.2$ and orbital period $P_{orb}=0.083333$ days ($120$ minutes).
The inner disc radius is derived using these parameters and assuming a
Keplerian velocity of $\sim2.37\times10^{3}$ km s$^{-1}$.
The 3:1 resonance radius is taken to be the outer disc radius.

\begin{myfigure}
\centerline{\resizebox{70mm}{!}{\includegraphics{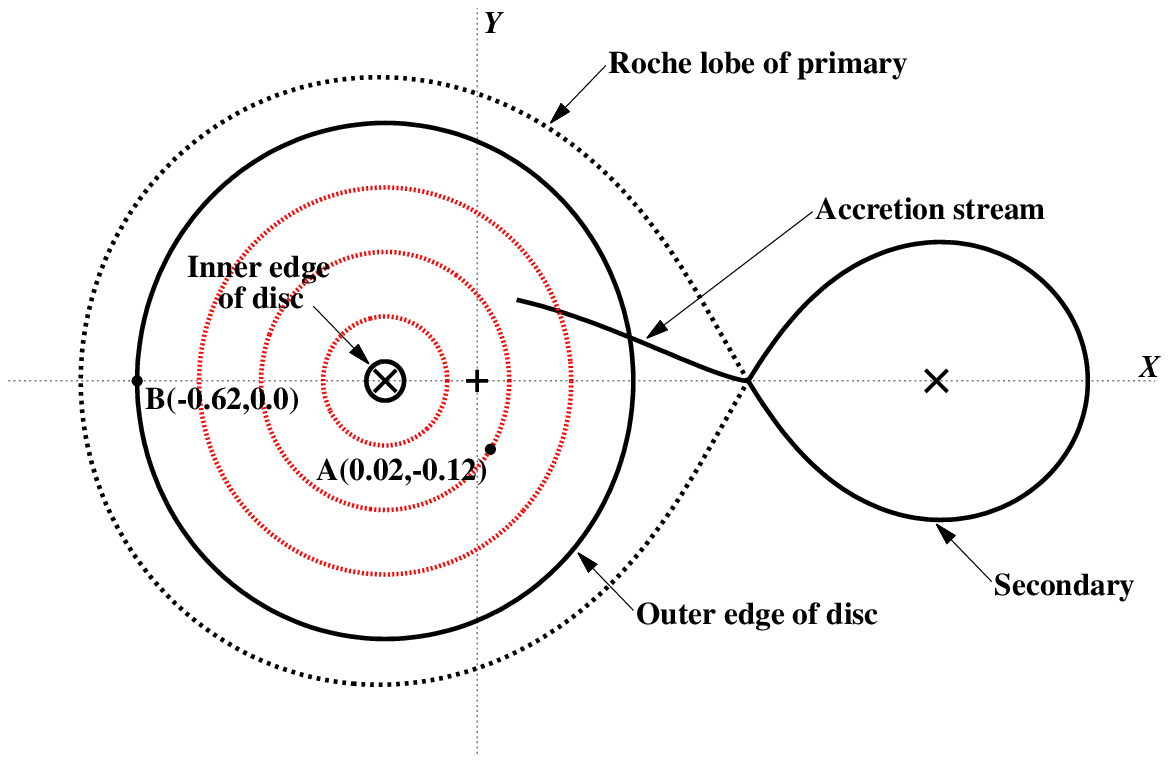}}}
\caption{Co-rotating Cartesian spatial coordinate frame for a model CV.}
\label{Enrico_fig01}
\end{myfigure}

\subsection{Velocity coordinates}

The left and middle panels of Fig.~2 show, respectively, the two-dimensional
Cartesian and polar velocity coordinate frames which correspond to the
co-rotating spatial frame shown in Fig.~1, with overlays of the velocity
profiles of all the main components of the model CV.
The polar velocity coordinate frame is obtained by either transforming the
Cartesian spatial frame to a polar spatial frame which is then projected into
a polar velocity frame or by transforming the Cartesian velocity frame into a
polar velocity frame.
Since we found that a polar frame allows for easier transformation of the
circularly symmetric velocity profile of a Doppler tomogram, this is the
first step in establishing an inside-out layout.

\end{multicols}
\begin{myfigure}
\centerline{\resizebox{160mm}{!}{\includegraphics{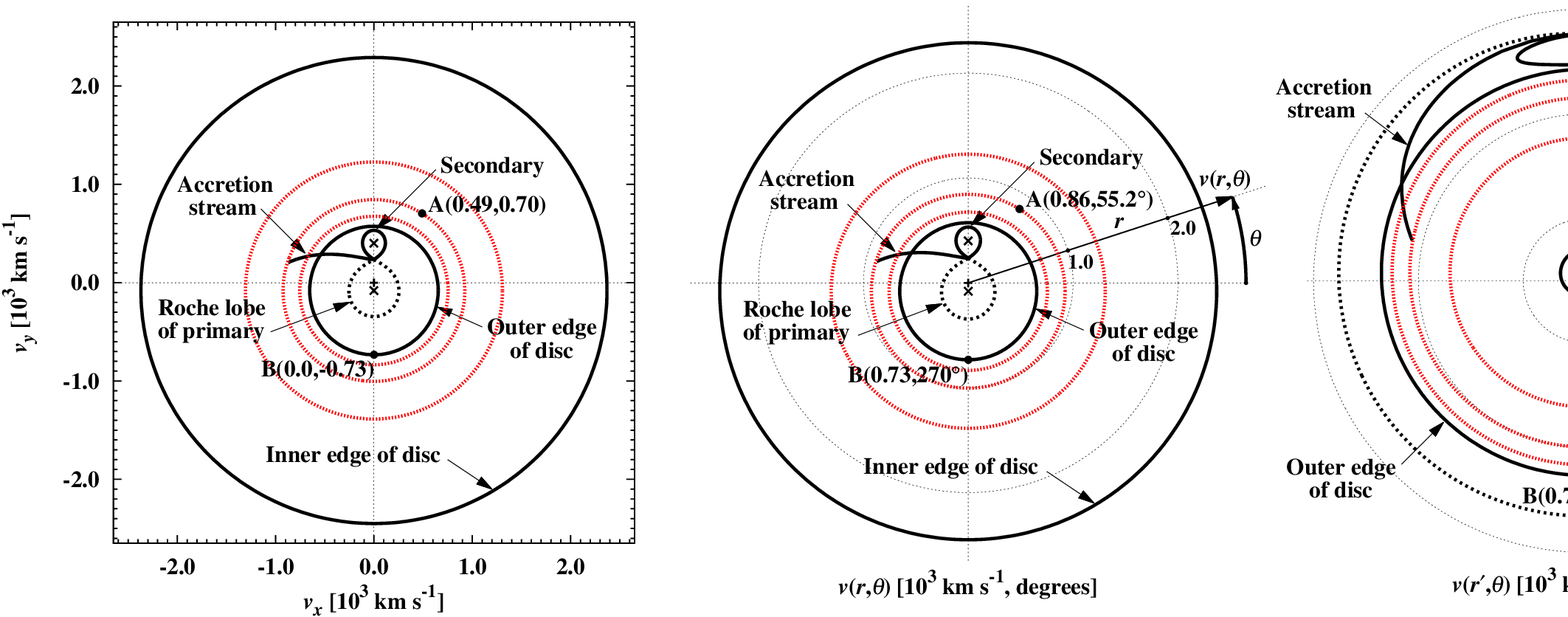}}}
\caption{Cartesian (left), polar (middle) and inside-out polar (right) velocity space.}
\label{Enrico_fig02}
\end{myfigure}
\begin{multicols}{2}

\section{Inside-out Doppler tomography}

The right panel of Fig.~2 shows the same model CV presented in the left and
middle panels, but with the zero velocity origin and the maximum velocities
transposed, creating an inside-out polar velocity space.
The most notable aspects of the inside-out frame are the inner and outer
edges of the accretion disc appearing the ``right'' way around while the
secondary and ballistic stream are outside the disc.
The secondary appears upside down because it is orbiting as a solid body,
i.e., the outside is moving faster than the inside.
The ballistic stream also ``curves'' inwards as it accelerates towards the
disc and primary.

\section{Examples}

The example Doppler tomograms have been created in polar and inside-out polar
velocity space using a modified version of the fast maximum entropy Doppler
mapping code presented by Spruit (1998).

\subsection{Spiral shocks in the accretion disc of the dwarf nova IP Peg}

Fig.~3 shows comparative non-axisymmetric normal and inside-out tomograms
for the HeII 4686\AA~emission line from phase-resolved spectroscopy of IP Peg.
The model velocity overlays were calculated using $P_{orb}=0.158206$ days
($\sim228$ minutes), $q=0.48$, $M_{1}=1.16$ and $i=83.8^{\circ}$
(Copperwheat et al.~2010).
The inner and outer edges of the disc (solid lines), the Roche lobes of the
primary (dashed line) and secondary (solid line) as well as a single particle
ballistic trajectory (solid line) from $L_{1}$ to $20^{\circ}$ in azimuth
around the primary, are shown.

In the normal tomogram the secondary appears as a bright spot inside the disc,
whereas it becomes a diffuse patch (spread over more pixels) outside the disc
in the inside-out tomogram.
However, in the inside-out tomogram the disc appears the ``right'' way around
as the two spiral shocks appear to be spiralling ``inward'' towards higher
velocities.

\subsection{Ballistic and magnetic accretion flow in the polar HU Aqr}

Comparative normal and inside-out tomograms for the HeII 4686\AA~emission
line from phase-resolved spectroscopy of HU Aqr are shown in Fig.~4.
$P_{orb}=0.086820$ days ($\sim125$ minutes), $q=0.4$, $M_{1}=0.875$ and
$i=84^{\circ}$ (one of the models from Schwope et al.~1997) were used to
calculate the model velocity overlays.
The Roche lobes of the primary (dashed line) and secondary (solid line) as
well as a single particle ballistic trajectory (solid line) from $L_{1}$ to
$65^{\circ}$ in azimuth around the primary, are shown.
A dipolar axis azimuth and co-latitude of $\sim 38^{\circ}$ and $\sim 12^{\circ}$
(Heerlein et al.~1999) respectively, were used to calculate magnetic dipole
trajectories (thin dotted lines) at $10^{\circ}$ intervals from $15^{\circ}$
to $65^{\circ}$ in azimuth around the primary.

The secondary appears as a bright spot in both the normal and the inside-out
tomograms.
In the normal tomogram the ballistic part of the accretion flow appears as a
prominent ridge with an apparent consistent brightness from $L_{1}$ to
$1.0\times10^{3}$ km s$^{-1}$, but with almost no discernible detail at
higher velocities.
In the inside-out tomogram the ballistic flow is more exposed and varying in
brightness, but retaining discernible detail to at least $1.5\times10^{3}$ km
s$^{-1}$.
Low-velocity ($0.0-0.5\times10^{3}$ km s$^{-1}$) emission associated with the
magnetic coupling region is seen as a diffuse patch in the lower left quadrant
of both tomograms.
There is no high-velocity ($>1.5\times10^{3}$ km s$^{-1}$) emission
discernible in the lower left quadrant of the normal tomogram, whereas in the
inside-out tomogram there is a small patch of emission between
$1.5-2.0\times10^{3}$ km s$^{-1}$ that can be linked to the magnetically
confined accretion flow close to the primary.

\section{Conclusions}

In a normal tomogram the lower-velocity features tend to dominate the
brightness scale since they are concentrated over fewer pixels compared to
higher-velocity features.
In an inside-out tomogram the converse is true with teneous higher-velocity
features being enhanced while prominent lower-velocity features are more
spread out and exposed.
Teneous lower-velocity features, however, may be diluted to the point of
becoming indiscernible similar to teneous higher-velocity features in a
normal tomogram.

We have shown that inside-out Doppler tomography projects the accretion disc
of a CV the ``right'' way around with the ballistic stream and the secondary
outside the disc.
Furthermore, the accretion flow of a polar appears more intuitive in an
inside-out tomogram with the ballistic stream curving ``inwards'' and the
magnetic flows being more exposed.
Therefore, we conclude that our new technique of inside-out Doppler tomography
is complementary to the existing technique.

\thanks
The authors would like to thank Axel Schwope and Danny Steeghs for sharing
their data of HU Aqr and IP Peg, respectively.

\end{multicols}
\begin{myfigure}
\centerline{\resizebox{140mm}{!}{\includegraphics{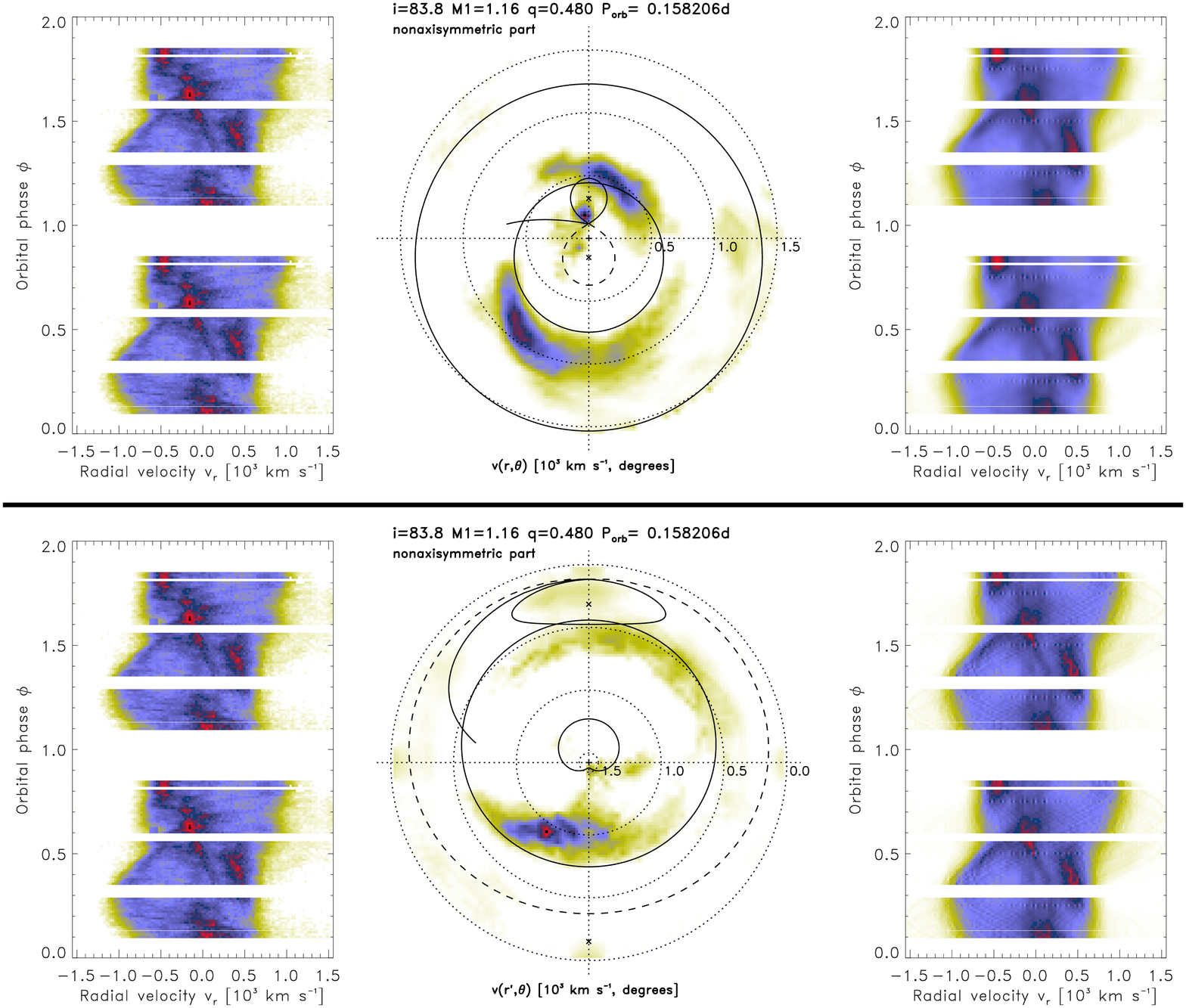}}}
\caption{Doppler tomography of IP Peg. Normal (top) and inside-out (bottom)
tomograms are shown for comparison.
The input and reconstructed trailed spectra are shown in the left and right
panels, respectively.}
\label{Enrico_fig03}
\end{myfigure}
\begin{multicols}{2}

\end{multicols}
\begin{myfigure}
\centerline{\resizebox{140mm}{!}{\includegraphics{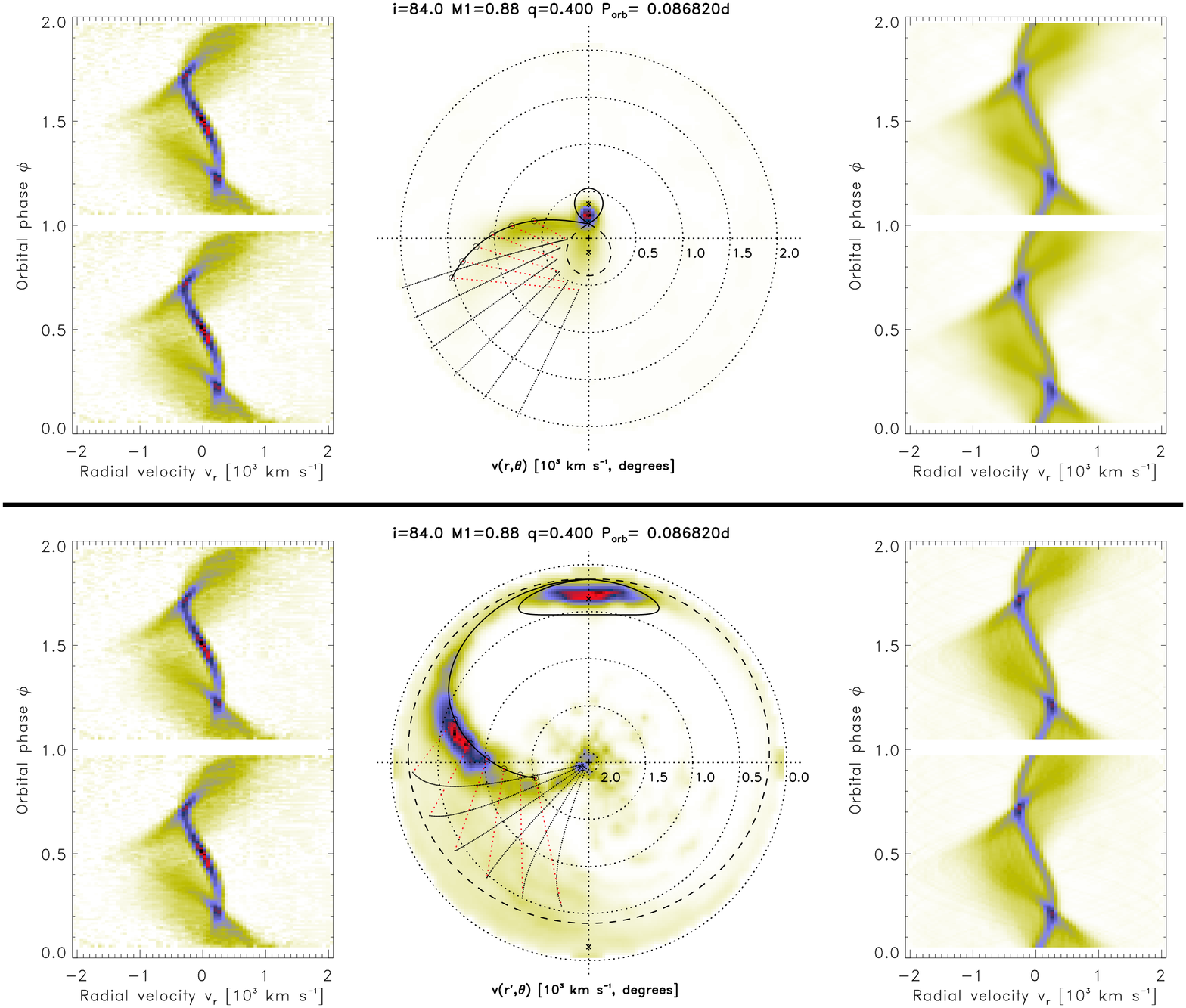}}}
\caption{Doppler tomography of HU Aqr. Normal (top) and inside-out (bottom)
tomograms are shown for comparison.
The input and reconstructed trailed spectra are shown in the left and right
panels, respectively.}
\label{Enrico_fig04}
\end{myfigure}
\begin{multicols}{2}

\bigskip
\bigskip
\noindent {\bf DISCUSSION}

\bigskip
\noindent {\bf DMITRY KONONOV:} How does your new technique redistributes
brightness and are there some physical suppositions behind this redistribution?

\bigskip
\noindent {\bf ENRICO KOTZE:} As we see in the example of IP Peg the upper
bright spiral shock in the normal tomogram appears less bright in the
inside-out tomogram since it is projected over a larger area of the image.
Similar to the normal technique the brightness distribution in the new
technique is purely a function of the projection of the spectra into the
velocity space frame and we have no claims that this is a representation of
the physical brightness distribution of the system.

\bigskip
\noindent {\bf LINDA SCHMIDTOBREICK:} Do you encounter problems with the
resolution as the low velocities are now spread over a large circle?

\bigskip
\noindent {\bf ENRICO KOTZE:} Yes, there can be a loss of resolution at low
velocities. This is the reverse of what happens in the normal tomograms where
high velocities are spread over a larger area and we can encounter a loss of
resolution in the high-velocity features.
That is why we feel the inside-out technique is complementary to the normal
technique.
Where the normal technique tends to enhance low-velocity features with some
loss in the resolution of high-velocity features, the inside-out technique
tends to enhance high-velocity features with some loss in the resolution of
low-velocity features.

\bigskip
\noindent {\bf DAVID BUCKLEY:} Is the magnetic longitude of the accreting
pole(s) a parameter in the inside-out maps?

\bigskip
\noindent {\bf ENRICO KOTZE:} Yes, for the model velocity profile overlays in
the inside-out tomograms of polars we take both the azimuthal and longitudinal
inclination of the assumed magnetic dipole into account.

\end{multicols}

\begin{thebibliography}{99}
\bibitem{} Copperwheat C.M., et al.: 2010, MNRAS, 402, 1824.
\bibitem{} Heerlein C., Horne K., Schwope A.D.: 1999, MNRAS, 304, 145.
\bibitem{} Marsh, T.R., Horne, K.: 1988, MNRAS, 235, 269.
\bibitem{} Schwope A.D., Mantel K.-H., Horne K.: 1997, A\&A, 319, 894.
\bibitem{} Spruit H.C.: 1998, arXiv:astro-ph/9806141.
\bibitem{} Warner B.: 1995, Cambridge Astrophysics Series 28, Cataclysmic
Variable Stars. Cambridge Univ. Press, Cambridge
\end{thebibliography}
\end{document}